\documentstyle[12pt]{article}

\def\newpic#1{%
   \def\emline##1##2##3##4##5##6{%
      \put(##1,##2){\special{em:point #1##3}}%
      \put(##4,##5){\special{em:point #1##6}}%
      \special{em:line #1##3,#1##6}}}
\newpic{}

\def\g{\gamma}
\def\G{\Gamma}

\def\sh{{\rm sh}}

\def\ctg{{\rm ctg}}

\def\r#1{\mbox{(}\ref{#1}\mbox{)}}

\def\te{\theta}
\def\e{{\rm e}}

\renewcommand{\theequation}{\arabic{section}.\arabic{equation}}

\def\b{\beta}
\def\a{\alpha}
\def\d{\delta}
\def\ep{\varepsilon}
\def\g{\gamma}

\def\h{{1\over 2}}
\def\sc{\scriptstyle}

\def\det{\mbox{det}}
\def\Sp{\mbox{Sp}}

\def\ba{\begin{array}}
\def\ea{\end{array}}
\def\beaa{\begin{eqnarray}}
\def\eeaa{\end{eqnarray}}
\def\bea*{\begin{eqnarray*}}
\def\eea*{\end{eqnarray*}}
\def\be{\begin{equation}}
\def\ee{\end{equation}}

\def\p{\psi}
\def\bp{\bar\psi}
\def\hd{\hat D}
\def\ha{\hat A}
\def\hb{\hat B}
\def\hk{\hat K}
\def\hp{\hat P}
\def\hj{\hat J}
\def\hg{\hat G}
\def\hf{\hat F}
\def\hee{\hat E}
\def\hl{\hat L}
\def\hc{\hat C}
\def\hr{\hat R}
\def\hq{\hat Q}
\def\hu{\hat U}
\def\hv{\hat V}
\def\hx{\hat \Xi}
\def\hi{\hat I}
\def\pro{\prod_{l=1}^{k}}
\def\prd{\prod_{l=1}^{2k}}
\def\xpx{{x,x^\prime}}
\def\const {\mbox{const}}
\def\px{x^\prime}
\def\oo {\mbox{O}}
\def\n{{(\nu)}}
\def\l{\left(}
\def\r{\right)}
\def\ro{\rho}
\def\o{\chi}

\begin{document}

\
\vspace{0.5cm}
\begin{flushright}
Preprint ITEP-TH-13/99\\
hep-th/9907040
\end{flushright}

\begin{center}
{\bf ASYMPTOTICS OF CORRELATION FUNCTION\\
OF  TWIST FIELDS IN \\
TWO DIMENSIONAL LATTICE FERMION MODEL}

\bigskip
{ A.I. Bugrij
\footnote
{e-mail:  abugrij@bitp.kiev.ua },
V.N. Shadura\footnote{e-mail:  shadura@bitp.kiev.ua }}

\medskip
 {\it Bogolyiubov Institute for Theoretical Physics}\\
 \medskip
 {\it 252143 Kiev-143, Ukraine}
\end{center}

\bigskip
\bigskip

\bigskip
\begin{abstract}
\begin{sloppypar}

 In  two-dimensional lattice fermion model a determinant representation
for the two-point correlation function of the twist field  in
the disorder phase is obtained. This field is defined by twisted boundary
conditions for lattice fermion field. The large distance asymptotics of
the correlation function
is calculated at the critical point and in the scaling region. The result
is compared with the vacuum expectation values of  exponential fields
in  the sine-Gordon model conjectured by S.Lukyanov and A.Zamolodchikov.
\end{sloppypar} \end{abstract} \thispagestyle{empty}

\renewcommand{\theequation}{1.\arabic{equation}}
\setcounter{equation}{0}

\bigskip
\centerline{\bf 1. Introduction}

\bigskip
During the last time  the progress has been made
in calculation of the long distance asymptotics
of a correlation functions of  local fields in some
integrable two-dimensional quantum field theories \cite{KS1,KO}.
Usually in these theories  two-point correlation functions
of  local fields can be represented as an infinite  series
of the form-factor contributions, which are calculated  using a method of
the form-factor bootstrap
\cite{Sm} or the angular quantization \cite{L1,LB}.
In
\cite{K,S}
a summation method of the form-factor decomposition
of the correlation function was developed. This method allows to obtain
a closed expression for the correlation function through
 Fredholm determinant of the integral operator
and to do  analysis of  asymptotic behaviour of one.

In this paper we analyse a large distance behaviour
of the two-point correlation function of the twist field
$\mu_\nu(r)$  in the disorder phase
of two-dimensional lattice fermion model.
At $\nu=\h$  this field coincides  with   the disorder field
$\mu (r)$  and the correlation function of these fields
satisfies the following relation \cite{ZI,ST}:
 \be
\langle\mu(0)\mu(r)\rangle=\langle\mu^I(0)\mu^I(r)\rangle^2 ,
\label {is}
\ee
where  $\mu^I(r)$  is  a disorder field in the Ising model.

Using the  functional integral method in
two-dimensional
lattice fermion model,
 we obtain a determinant representation
of the  correlation function.
Calculating  the asymptotics of the correlation function
 for $r\to\infty$,
 we find the vacuum expectation values
$\langle \mu_\nu(r)\rangle $.
Here index $\nu$ ($0<\nu <1$) denotes  twisted boundary conditions
for lattice
fermion field  along the correlation line. In scaling limit
(massive  fermion
quantum field theory) this field is massive analog
of the twist field  in two dimensional conformal field theory
with central charge $c=1$ \cite{GI}.

In the free fermion point of
 the sine-Gordon model the vacuum expectation value\hfill\break
 $<\mu (r)>$ is connected by simple relation with
 the vacuum expectation value
of the exponential field
$\exp(i\nu\phi) $ for $\nu=\h$.
The sine-Gordon model
is described by the Euclidean action
$$
S={1\over{4\pi}}\int dxd\tau \left(\partial_k\phi\partial_k\phi+
\mu\cos(\hat\b\phi) \right).$$
The free fermion point occurs for $\hat\b=1$.

In this point the exponential field can be defined by the
braiding relations with the free fermion field
\cite{SJM}:
$$
\e^{i\nu\phi(\tau, x)}\psi(\tau, y)=\psi(\tau, y)\e^{i\nu\phi(\tau, x)},
\quad\mbox{if\ } y<x,$$
\be \e^{i\nu\phi(\tau, x)}\psi(\tau, y)=\e^{2\pi
i\nu}\psi(\tau, y)\e^{i\nu\phi(\tau, x)},  \quad\mbox{if\ } y>x.
\label {ex}
\ee
As it was shown in
\cite{ZI,ST,BL} the correlation functions of the  disorder fields
  satisfy the following relations
\be
\langle \mu(0)\mu(r)\rangle  =
\langle \cos {\sc\h}\phi(0)\, \cos {\sc\h}\phi(r) \rangle .
\label {is1}
\ee
 From here  one gets
\be
\langle\mu(0)\mu(r)\rangle>=
\h\langle\e^{i\h\phi(0)}\,\e^{i\h\phi(r)}\rangle
+\h\langle\e^{i\h\phi(0)}\,\e^{-i\h\phi(r)}\rangle.
\label{kors}
\ee
Using  this  equation, we can obtain   a relation between
the vacuum expectation values of the disorder and exponential  fields.

Note that for calculation of the left and right hand sides
of  (\ref{kors})
it is nessesary to regularize  explicit expressions for
the correlation functions that it leads to appearence of
the corresponding wave  renormalization constants.
To be rid of them at calculation of the expectation values
it is convenient to normalize the correlation functions
on behaviour for $r\to 0$. In this limit we obtain from  (\ref{kors})
\be
\langle\mu(0)\mu(r)\rangle=
\h\langle\e^{i\h\phi(0)}\,\e^{-i\h\phi(r)}\rangle,
\label{kors1}
\ee
where  we have used the following asymptotics for the correlation function
of the exponential fields for $r\to 0$ \cite{LZ}
\be
\langle\e^{i\nu\phi(0)}\,\e^{i\nu^{\prime}\phi(r)}\rangle=
\langle\e^{i(\nu+\nu^{\prime})\phi}\rangle r^{2\nu\nu^{\prime}}.
\label{as}
\ee
Then for a ratio  of
 (\ref{kors}) in the limit  $r\to  \infty$ and
(\ref{kors1}) one gets
\be
\frac{\langle\mu(0)\mu(r)\rangle>|_{r\to
\infty}}{\langle\mu(0)\mu(r)\rangle|_{r\to 0}}=
\frac{\langle\mu\rangle^2}{\langle\mu(0)\mu(r)\rangle|_{r\to 0}}=
\frac{2\langle\e^{i\h\phi}\rangle^2}{\langle\e^{i\h\phi(0)}\,
\e^{-i\h\phi(r)}\rangle|_{r\to 0}}.
\label{kors2}
\ee
As it will shown below one can choose such normalization of the disorder
fields that asymptotic behaviour of the correlation functions in
denominators of the left and right hand sides of  (\ref{kors2})
coincides.
In this case we obtain from (\ref{kors2})
$\langle\mu\rangle=\sqrt 2 \langle\exp(i\h \phi)\rangle$.
In the paper
\cite{LZ} (also see \cite{P}) it was conjectured
 the following expression for the
vacuum expectation values   of the exponential fields ($|\nu|<1$)
 \be
\langle\e^{i\nu\phi}\rangle= \left(\frac{m}{2}\right)^{\nu^2}
\exp\left[\int_{0}
^{\infty}{{dt}\over t}\left({{\sh^2{{\nu t}\over 2}}
\over{\sh^2{t\over 2}}}
-\nu^2\e^{-t}\right)\right].
\label{asy6}
\ee
This  expression it is easy to obtain by means of
results of the paper \cite{T2}. Here
the  asymptotic behaviour of the correlation function of the exponential
fields for $r\to 0$ was calculated.
Comparing  our result (\ref{asy5}) and
(\ref{asy6}), we obtain
\be
\langle\mu_\nu\rangle=4^{\nu^2} \langle\exp(i\nu \phi)\rangle.
\label{vev}
\ee
 Emphasize that this relation is correct for such  normalization
of the fields
that the correlation functions of the twist and exponential fields
have  the same asymptotics for $r\to 0$
$$
\langle\mu_\nu(0)\,\mu_\nu(r)\rangle
\mathop{=}_{r\to 0}<\e^{i\nu\phi(0)}\,\e^{-i\nu\phi(r)}>
\mathop{=}_{r\to 0}\frac{1}{r^{\,2\nu^2}}.
$$

In section 2
the determinant
representation of the two-point correlation function  of the twist field
is  derived.  In section 3
asymptotic  behaviour of the correlation function
are considered and  new determinant representation through
the matrix of the
Toeplitz type are obtained in the limit $r\to\infty$.
In section 4, using this
representation, asymptotics of the correlation function is
calculated. In Appendices some relations used in this
paper are derived.

\renewcommand{\theequation}{2.\arabic{equation}}
\setcounter{equation}{0}

\bigskip
\begin{center}
{\bf 2.  Functional integral representation \\
of the  correlation function }

\end{center}

\bigskip

It is known
\cite{MS2,BS},
that the correlation function
 $\langle \mu^I(0)\mu^I(r)\rangle $
in the two dimensional Ising model can be represented
in the form of the   functional integral for the lattice Majorana
fermion  theory
with  antiperiodic boundary conditions for the fermion field with the
exception of the line $[0,r-1]$, where the  fermion field has periodic
boundary conditions.

The same way let us define the two-point correlation function of
  the field
$\mu_\nu(r)$   in the lattice Dirac fermion  theory
through a ratio of the following functional integrals
\be
\langle \mu_\nu(0)\mu_\nu(r)\rangle ={{\int d[\psi\bp]
\e^{S_r[\p]}}\over
{\int d[\psi\bp]\e^{S[\p]}}},
\label {cor}\ee
with  the action
\be
S_r[\p]=(\bp\hd\p)= \sum_{\ro,\ro^\prime}\bp (\ro)D_{\ro,\ro^\prime}
\p(\ro^\prime), \label {act}
\ee
where $\p(\ro)$  is a complex two component
grassmann field; the coodinates of the lattice sites
$\ro=(x,y)$ run through the values  $x=1, ..., n_x$,
$y=1, ..., n_y$; $\hd$
is the lattice Dirac operator
\be
\hd=\left(\begin{array}{cc} u&v\\ -v^T&u^T\end{array}
\right)= \left( \begin{array}{cc} 1-t\nabla_x&1-t\nabla_y \\
-(1-t\nabla_{-y})&1-t\nabla_{-x} \end{array} \right).
\ee
Here
 $\nabla_x$, $\nabla_y$ denote the shift  operators along
the  $X$ and  $Y$ axes:
$$
\nabla_x \p(\ro)=\p(\ro+\hat x),\quad \nabla_y \p(\ro)=
\p(\ro+\hat y),
$$
where $\hat x$,  $\hat y$  are unit vectors.
Index ``$r$"  denotes  that in action
  $S_r[\p]$ operator $\nabla_y$ has  twisted boundary conditions
along the line $[0,r-1]$ ($r-1=(r-1,0)$):
$$
\nabla_y \p(x,n_y)=-\e^{2\pi i\nu}\p(x,0),\quad
x=0,1,\ldots,r-1.
$$
In denominator
(\ref {cor})  the action $S[\p]$  coincides
with $S_r[\p]$ by the form, but $\nabla_y$ satisfies
 antiperiodic boundary
conditions along all boundary.

It is not hard to check that the operator
  $\hd$  satisfies the  relation
 $\hd^{-1}=\hd^T\cdot\hk$,
 where $\hk$  is the lattice Klein-Gordon  operator
$$
\hk=\det \hd =u u^T+v v^T= 2(1-t)^2 -t
[(\nabla_x-2+\nabla_{-x}) +(\nabla_y-2+\nabla_{-y})].
$$
In the "naive"  continuum limit (the lattice constant
$a \rightarrow 0$)
$\hk=a^2 t(m^2 -\partial_i\partial_i)$, where
  $\partial_i=(\nabla_i-1)/a$
 and $ma/\sqrt 2 =t^{-\h}-t^{\h}$.
From here   the critical point of the model  is
 $t_c=1$,  and the scaling region is determined by relation
 $|t_c-t|=ma/\sqrt 2$.
In momentum representation  the operator $\hat K$ is diagonal and
 the function
$$ K(p)=2(1-t)^2+4t\biggl(\sin^2{{p_x}\over 2}+\sin^2{{p_y}\over
2}\biggr)
$$
  has unique minimum at
$p_x=p_y=0$ in the Brillouin zone and
unlike the usual lattice Dirac operator
 we have  not problem with the fermion dubling.

In order to integrate over the fermion fields in the numerator
(\ref{cor}), in the action $S_r[\p]$ it is convenient  to go
 to operator  $\nabla_y$ with antiperiodic boundary conditions
along  all boundary. After that  we get
 additional term  $\delta S_r[\p]$
in the
action $S_r[\p]$, which contains a lattice defect:
$$
 S_r[\p]=S[\p]+\d S_r[\p],
$$
where $$ \d
S_r[\p]=t\sum_{x=0}^{r-1}[-\xi\bp^1(x,n_y)\p^2(x,0)+
\xi^*\bp^2(x,0)\p^1 (x,n_y)]=(\bp \hp^T\hx\hp\p).
$$
When
\be \langle \mu_\nu(0)\mu_\nu(r)\rangle =\langle \e^{\d S_r[\p]}
\rangle =
{{\int d[\psi\bp]\e^{S[\p]+\d S_r[\p]}}\over {\int
d[\psi\bp]\e^{S[\p]}}}. \label {cor1}
\ee
Here the projective operator  $\hp$ selects  lattice sites on the line
$[0,r-1]$
$$
\hp_{l,x}=\d_{l,x}\d_{n_y,y}\left(\begin{array}{cc} 1&0\\ 0&\nabla_y
\end{array} \right), \quad l=0,1,\ldots,r-1; $$
  and
$$
\hx=t\left(\begin{array}{cc} 0&-\xi \hi^{(r)}\\ \xi^*\hi^{(r)}&0
\end{array} \right),$$ where $\xi=1-\e^{2\pi i\nu}$
and $\hi^{(r)}$  is a unit matrix of the size
  $r\times r$.

For calculation of the functional integrals in
 (\ref{cor1}) it is convenient to represent
$\e^{\d S_r[\p]}$  through the integral over a auxiliary field
\be
\e^{\d S_r[\p]}= |\hx|\int d[\o\bar\o] \exp[{(\bar\o\hx^{-1}\o)
+(\bar\o\hp\p)+(\bp\hp^T\o)}].
 \ee
 Using(2.5) and integrating  over  $\p(\ro)$  in
 (\ref{cor1}), one gets
\beaa
\langle \mu_\nu(0)\mu_\nu(r)\rangle
\!\!&=&\!\!
|\hx|\int d[\o\bar\o] \exp
[{(\bar\o\hx^{-1}\o)+(\bar\o\hp\hd^{-1}\hp^{T}\o)}]=\nonumber\\
\!\!&=&\!\!
|\hx|\cdot |\hx^{-1}+\hp\hd^{-1}\hp^T|=|\hg|,\label{cor2}
\eeaa
where $|\hg|$ denotes the determinat of a block matrix of the dimension
  $(2\times 2)$ with blocks of the dimension
  $(r\times r)$. The matrix $\hg$  has the form
\be
\hg=\left(\begin{array}{cc}g^{11}_{x,x^\prime}\sin\pi\nu&
-g^{12}_{x,x^\prime}\sin\pi\nu+iI_{x,x^\prime}^{(r)}\cos\pi\nu\\
 g^{12}_{x,x^\prime}\sin\pi\nu +iI_{x,x^\prime}^{(r)}\cos\pi\nu&
          (g^{11}_{x,x^\prime})^T\sin\pi\nu
\end{array} \right),\label{mg}\ee
where
$$
g^{11}_{x,x^\prime}={1\over{(2\pi)^2}}\int
\limits_{-\pi}^{\pi}{{d^2p}\over{K(p)}}
\e^{ip_x(x-x^\prime)}2t u^*(p_x),\quad u(p_x)=1-t\e^{ip_x},$$
$$
g^{12}_{x,x^\prime}={1\over{(2\pi)^2}}\int\limits_{-\pi}^{\pi}{{d^2p}
\over{K(p)}}\e^{ip_x(x-x^\prime)}2t(u^*(p_x)+u(p_x)).
 $$
Let us  show that the correlation function
 (\ref {cor2}) for $\nu=\h$  is equal to  square of
 the correlation function of the disorder field  in the Ising model.
 For this  it is necessary  to do
 a similarity
transformation of the matrix $\hg$ in (\ref {cor2})
by means of the unitarity matrix $\hq$
 \be
\hq=\e^{i\gamma_5{\te\over 2}}, \quad \g_5= \left(\begin{array}{cc}
0&-i\\ i&0 \end{array} \right),\quad  \ctg\te=t,\label {tr}
\ee
As result we obtain
\be
G(r)=\langle \mu_\nu(0)\mu_\nu(r)\rangle
\equiv|\hq\cdot\hg\cdot\hq^{-1}|=|\cos^2\pi\nu +\sin^2\pi\nu
\hv\cdot\hv^T|,\label {cor3}\ee
where
 \be V_{x,x^\prime}={1\over{2\pi
i}}\oint\limits_{|z|=1}{{dz}\over{z}} z^{-(x-x^\prime)}V(z), \quad
V(z)=\sqrt{{(1-\a z)(1-\b z^{-1})}\over{(1-\a z^{-1})(1-\b z)}},
\label{kor}\ee\be \a=t{{\sqrt{1+t^2}+t}\over{\sqrt{1+t^2}+1}},
\quad\b=t{{\sqrt{1+t^2}-t}\over {\sqrt{1+t^2}+1}}.\label{par}
\ee
Here the matrix
 $V_{x,x^\prime}$ exactly coincides with classical expression
for the Toeplitz matrix  \cite{MW}, the determinant of which determines
the correlation function
$\langle \mu^I(0)\mu^I(r )\rangle $
in the paramagnetic phase of  the Ising model.
It is obvious that
(\ref{is}) follows from (\ref{cor3}) for $\nu=\h$.
In further we  include
  coefficient $(2\pi i)^{-1}$  in the integration measure of
 contour integrals.

Note that the transformation
$\hq$   is connected with particular version of the lattice
Dirac operator
(2.3).
In the "naive" continuum limit  we obtain for the action (2.2)
\be S[\p]=\int d^2 \ro \bp(\ro)(m\e^{i\g_5{\pi\over
4}}-\g_i\partial_i)\p(\ro),
\ee
where
$$ \g_x=\left(\begin{array}{cc} 1&0\\
0&-1\end{array} \right),\quad \g_y= \left(\begin{array}{cc} 0&-i\\
i&0\end{array} \right).
$$
At the critical point $t_c=1$ the angle $\te$ in
(\ref{tr}) is equal to $\pi/4$  and the transformation $\hq$ is
the $\gamma_5$-rotation of the Grassmann field:  $\p^\prime
 =\exp(i\g_5\pi/8)\p$,
after that the Dirac action takes the usual form
\be
S[\p]=\int d^2 \ro \bp(\ro)(m-\g_i\partial_i)\p(\ro).
\ee

\renewcommand{\theequation}{3.\arabic{equation}}
\setcounter{equation}{0}

\bigskip

\begin{center}
{\bf 3.
  Toeplitz determinant representation
\\ of the correlation function}
\end{center}

\bigskip

In this section   we show that in the scaling limit  ($ ma\to 0$)
at $r\to \infty$ a evaluation of the correlation function (\ref{cor3})
can be reduced
to  a calculation of the Toeplitz determinant:
\be
G(r)=|\cos^2\pi\nu +\sin^2\pi\nu \hv\cdot\hv^T|\mathop{=}_{r\to\infty}
|V^{(\nu)}\cdot V^{(\nu)^T}|=|V^{(\nu)}|^2,\label {cor4}\ee
where
\be
V^{(\nu)}_{x,x^\prime}=\oint\limits_{|z|=1}{{dz}\over{z}}
z^{-(x-x^\prime)}V^{(\nu)}(z),\quad V^{(\nu)}(z)=\left[{{(1-\a z)(1-\b
z^{-1})}\over{(1-\a z^{-1})(1-\b z)}}\right] ^\nu
\label{kon}\ee
is a Toeplitz matrix.
For $\nu=\h$ the kernel $V^{(\nu)}(z)$  coincides with  the kernel $V(z)$
(\ref{kor}).

In Appendix A it is shown that the correlation function
  (\ref {cor3}) can be represented in the following form
\be G(r)\mathop{=}_{r\to\infty}|1-\sin^2\pi\nu\ha|^2,\label
{fa1}\ee
where
 $$
A_\xpx =\oint\limits_{|z_1z_2|<1} {{dz_1dz_2 (z_1)^x
(z_2)^{\px}}\over{(1-z_1z_2)}} V^{-1}(z_1)V^{-1}(z_2).
$$
Here the integration contour passes between the cuts  depicted
on Fig.~1.

\special{em:linewidth 1.0pt}
\unitlength 1.00mm
\linethickness{0.3pt}
\begin{picture}(50.00,115.00)(0.00,15.00)
%\circle(60.00,80.00){60.67}
\emline{60.00}{110.34}{1}{64.85}{109.95}{2}
\emline{64.85}{109.95}{3}{69.57}{108.79}{4}
\emline{69.57}{108.79}{5}{74.04}{106.89}{6}
\emline{74.04}{106.89}{7}{78.16}{104.30}{8}
\emline{78.16}{104.30}{9}{81.81}{101.09}{10}
\emline{81.81}{101.09}{11}{84.89}{97.34}{12}
\emline{84.89}{97.34}{13}{87.34}{93.14}{14}
\emline{87.34}{93.14}{15}{89.09}{88.60}{16}
\emline{89.09}{88.60}{17}{90.09}{83.84}{18}
\emline{90.09}{83.84}{19}{90.32}{78.99}{20}
\emline{90.32}{78.99}{21}{89.77}{74.16}{22}
\emline{89.77}{74.16}{23}{88.45}{69.48}{24}
\emline{88.45}{69.48}{25}{86.41}{65.07}{26}
\emline{86.41}{65.07}{27}{83.68}{61.04}{28}
\emline{83.68}{61.04}{29}{80.35}{57.50}{30}
\emline{80.35}{57.50}{31}{76.49}{54.54}{32}
\emline{76.49}{54.54}{33}{72.22}{52.23}{34}
\emline{72.22}{52.23}{35}{67.62}{50.64}{36}
\emline{67.62}{50.64}{37}{62.84}{49.80}{38}
\emline{62.84}{49.80}{39}{57.98}{49.73}{40}
\emline{57.98}{49.73}{41}{53.17}{50.44}{42}
\emline{53.17}{50.44}{43}{48.53}{51.92}{44}
\emline{48.53}{51.92}{45}{44.20}{54.11}{46}
\emline{44.20}{54.11}{47}{40.26}{56.96}{48}
\emline{40.26}{56.96}{49}{36.84}{60.41}{50}
\emline{36.84}{60.41}{51}{34.01}{64.36}{52}
\emline{34.01}{64.36}{53}{31.84}{68.72}{54}
\emline{31.84}{68.72}{55}{30.40}{73.36}{56}
\emline{30.40}{73.36}{57}{29.72}{78.17}{58}
\emline{29.72}{78.17}{59}{29.82}{83.03}{60}
\emline{29.82}{83.03}{61}{30.69}{87.82}{62}
\emline{30.69}{87.82}{63}{32.31}{92.40}{64}
\emline{32.31}{92.40}{65}{34.65}{96.66}{66}
\emline{34.65}{96.66}{67}{37.64}{100.50}{68}
\emline{37.64}{100.50}{69}{41.20}{103.80}{70}
\emline{41.20}{103.80}{71}{45.24}{106.50}{72}
\emline{45.24}{106.50}{73}{49.66}{108.52}{74}
\emline{49.66}{108.52}{75}{54.35}{109.80}{76}
\emline{54.35}{109.80}{77}{60.00}{110.34}{78}
%\end
\special{em:linewidth 0.5pt}
\put(70.33,74.67){\makebox(0,0)[ct]{$\beta$}}
\put(85.33,74.67){\makebox(0,0)[ct]{$\alpha$}}
\put(107.33,74.67){\makebox(0,0)[ct]{$\alpha^{-1}$}}
\put(134.67,74.00){\makebox(0,0)[ct]{$\beta^{-1}$}}
\put(74.33,29.00){\makebox(0,0)[cc]{$ Fig. 1.$}}
\put(24.67,82.67){\makebox(0,0)[cc]{$-1$}}
\put(91.67,82.67){\makebox(0,0)[cc]{$1$}}
\put(113.67,115.00){\makebox(0,0)[cc]
{z}}
%{\large z}}
\put(70.00,79.33){\rule{15.00\unitlength}{1.00\unitlength}}
\emline{69.67}{81.00}{79}{69.67}{78.33}{80}
\emline{85.33}{81.00}{81}{85.33}{78.33}{82}
\put(107.33,79.33){\rule{27.33\unitlength}{1.00\unitlength}}
\emline{107.00}{81.00}{83}{107.00}{78.33}{84}
\emline{135.00}{81.00}{85}{135.00}{78.33}{86}
\emline{109.33}{119.33}{87}{109.33}{111.33}{88}
\emline{109.33}{111.33}{89}{119.33}{111.33}{90}
\put(58.00,82.67){\makebox(0,0)[cc]{$0$}}
\emline{15.00}{80.00}{91}{150.00}{80.00}{92}
\emline{60.00}{40.00}{93}{60.00}{125.00}{94}
\put(77.50,79.50){\oval(19.00,5.00)[]}
\special{em:linewidth 0.2pt}
\emline{79.00}{82.50}{95}{76.00}{82.00}{96}
\emline{76.00}{82.00}{97}{79.00}{81.50}{98}
\emline{85.00}{95.00}{99}{87.66}{93.00}{100}
\emline{85.00}{95.00}{101}{85.00}{92.00}{102}
\end{picture}

Using (3.3),
we get
\be
{\sc\h}\ln G(r)=\Sp\ln(1-\sin^2\pi\nu  \ha)=-\sum_{k=1}^{\infty}{a_k
\over k} \left({{\sin\pi\nu}\over\pi}\right)^{2k},
\label{cor5}\ee
where
\be a_k =\pi^{2k}\Sp \ha^k=\pi^{2k}\oint \prd dz_l
 {{\pro[1-(z_{l}z_{l+1})^r]} \over{\prd[(1-z_lz_{l+1})V(z_l)]}}, \quad
z_{2l+1}=z_1.\label {ko1}\ee
For  $r\to \infty$ the terms  in the right-hand side of
(\ref {ko1}), which contain the product $(z_lz_{l+1})^r$,
are exponentially small, so that
\be a_k =\pi^{2k}\oint{{\prd
dz_l}\over{\prd[(1-z_lz_{l+1})V(z_l)]}}= \int\limits_{\b}^{\a}{{\prd
dz_l}\over{\prd(1-z_lz_{l+1})}}\prd\left[{{(\a- z_l) (1-\b
z_{l})}\over{(1-\a z_{l})(z_l-\b)}}\right ]^\h.\label {ko2}
\ee
 At
$ma=0$ ($\a=1$) the integral in (\ref{ko2}) has logarithmic divergence
  on the upper limit (in  a neighborhood of the  "upper right cone"
of the $2k$-dimensional hypercube with the coordinates
$z_l=1$).
Let us isolate  and evaluate this divergence:
\be
a_k\simeq\int\limits_{\a-\ep}^{\a}\prd{{ dz_l}\over{1-z_lz_{l+1}}}\prd
\left({{\a-z_l} \over{1-\a z_l}}\right)^\h{\simeq}
\int\limits_{1}^{\ep/ma}\prd {{dz_l}\over{z_l+z_{l+1}}}.\label{cep}
 \ee
For  derivation of this estimate we did  the following substitutions of
the integration variables:
$z_l=\a-z^\prime_l,$ $z^\prime_l=(ma)z^{\prime\prime}_l,$
$z^{\prime\prime}_l=z_l+1$.
Isolating the logarithmic singularity in the last integral,
we obtain
\be
a_k=\int\limits_{1}^{\ep/ma} \prd {{dz_l}\over{z_l+z_{l+1}}}=-
kb_k\ln(ma)+\const.
\label{ko3}\ee
Differentiating over
$ma$ the left- and right-hand sides of (\ref {ko3}),
the coefficients $b_k$ can  be expressed   through the multiple integrals
\be
b_k =2\int\limits_{0}^{1}\prod_{l=1}^{2k-1}dz_l\cdot
\biggl[(1+z_1)(1+z_{2k-1})
\prod_{l=1}^{2k-2}(z_l+z_{l+
1})\biggr]^{-1}.
\label{ko4}\ee
Using
 (\ref{ko3}) and (\ref{cor5}),
we get  the asymptotic evaluation of the correlation
function for $ r\to\infty$, $ma\to 0$
\be
{\sc\h}\ln G(\infty)=
\ln(ma)\sum_{k=1}^{\infty}b_k\left({{\sin\pi\nu}\over
{\pi}}\right)^{2k} + \const.\label{cor6}
\ee
The integrals in the first three coefficients in
 (\ref {ko4})  can be reduced to the table ones
\be
b_1 =1, \quad b_2 =2\zeta(2),\quad b_3=
4\zeta^2(2)+6\zeta(4),\label{ko5}\ee
where $\zeta(k)$ is the Rieman $\zeta$-function
$$
\zeta(k+1)={{(-1)^k}\over{k!}}\int\limits_{0}^{1}{{dx\ln^k x}\over{1-x}}.
$$
Decomposing in  (\ref {cor6}) the functions
 $(\sin\pi\nu/\pi)^{2k}$
over powers of  $\nu$  and using the values (\ref {ko5}) for
the coefficients $b_k$, it is not difficult to get
$$
b_1
{{\sin^2\pi\nu}\over{\pi^2}}=\nu^2 +\oo (\nu^4),$$ \be b_1
{{\sin^2\pi\nu}\over{\pi^2}}+b_2 {{\sin^4\pi\nu}\over{\pi^4}} =
\nu^2 +\oo
(\nu^6),\label{ko6}\ee $$ b_1 {{\sin^2\pi\nu}\over{\pi^2}}+b_2
{{\sin^4\pi\nu}\over{\pi^4}}+ b_3 {{\sin^6\pi\nu}\over{\pi^6}}=
\nu^2 +\oo
(\nu^8),
$$
i.e. every following  term in the series
 (\ref {ko5}) annihilates   more high powers
of  $\nu^{2k}$.

Although the integrals
(\ref{ko4})  is hardly calculated for multiplicity higher five,
one can assume that in more high  orders relations of the type
(\ref {ko6}) also take place. Then
 \be
\sum_{k=1}^{\infty}b_k\left({{\sin\pi\nu}\over{\pi}}\right)^{2k}=
\nu^{2}.
\ee
It means that the generation function for the coefficients
 $b_k$ has the form
\be
{1\over{\pi^2}}\arcsin^2(\pi\sqrt{z})=\sum_{k=1}^{\infty}b_k z^k.
\label{fu}
 \ee
Then  for the determinant
 (\ref {cor4}) we obtain
 \be
G(\infty)\mathop{=}_{(ma)\to
0}\const\cdot(ma)^{2\nu^2}.\label{asco}
\ee
Note that  (\ref {fu}) allows to find values of the integrals
(\ref{ko4}) of  arbitrary multiplicity.

Let us prove the asymptotics (\ref{asco}).
The following observation is  key for the proof (it follows from the
calculations in (\ref{cep})):
the asymptotic evaluation ($ma\to 0$) of  the integrals in
 (\ref{ko2}) is insensitive to the exponent in the expression for
kernel of
the matrix $V_{\xpx}$.  It allows  to introduce the new
matrix $ V^{(\nu)}_{\xpx}$ with the kernel (\ref{kon}).
  Let us  define the
correlation function \be G^\n(r)=|\hv^\n\cdot \hv^{\n\,\,T}|,
\label{cn1} \ee
where the matrix $\hv^\n$ is determinated in (\ref{kon}).
Using the results of Appendix A,
(\ref{cn1}) can be represented  as
\be
G^\n(r)\mathop{=}_{r\to\infty}|1-\ha^\n|^2,\label{cn2}\ee
where the matrix $\ha^\n$  has the form
 \be
A^\n_\xpx =\oint {{dz_1dz_2 (z_1)^x (z_2)^{\px}}
\over{ (1-z_1z_2)}}{1\over{V^{\n}(z_1)V^{\n}(z_2)}}.\label{mn1}
\ee
The trace of the
 $k$-th power of the  matrix  $\ha^\n$
is expressed through
 the $2k$-multiple integral
\be
\Sp[(\ha^\n)^k]= \left({{\sin\pi\nu}\over{\pi}}\right)^{2k}
\int\limits_{\b}^{\a}\prd
{{ dz_l}\over{(1-z_lz_{l+1})}}\prd
\left[{{(\a- z_l)(1-\b z_{l})}\over{(1-
\a z_{l})(z_l-\b)}}\right ]^\nu\prod_{l=1}^{k}
[1-(z_lz_{l+1})^r].\label{sp1}
\ee
Here, unlike
 (\ref{cor5}), the factor $(\sin\pi\nu)^{2k}$
 appeares on account of the exponent $\nu$
at contraction of the integration contour.
Taking into account (\ref{sp1}) for $r\to\infty$,
 we get the expression for $G^\n(r)$
which is similar to (\ref{cor5})
\be
{\sc\h}\ln G^\n(r)=\Sp\ln(1- \ha^\n)=-\sum_{k=1}^{\infty}{a^\n_k
\over
k}\left({ {\sin\pi\nu}\over\pi}\right)^{2k},\label{cor7}\ee
where
\be a^\n_k
=\int\limits_{\b}^{\a}\prd{{ dz_l}\over{(1-z_lz_{l+1})}}\prd\left[
 {{(\a-
z_l)(1-\b z_{l})}\over{(1-\a z_{l})(z_l-\b)}}\right ]^\nu
\prod_{l=1}^{k}[1-(z_lz_{l+1})^r].\label {ko7}
\ee
For $r\to\infty$ the terms in (\ref{ko7}) containing the power of
$(z_lz_{l+1})^r$ disappear. Although the rest integrals in (\ref{ko7})
 diverge on the upper limit the difference between (\ref{ko2}) and
(\ref{ko7})
\beaa \!\!\!\!\!\Delta a_k^\n\!\!& \equiv &\!\!
a_k-a_k^\n=\nonumber \\ \!\!&=\!\!&\int\limits_{\b}^{\a}\prd{{
dz_l}\over{(1-z_lz_{l+1})}}\left\{\prd\left[{{(\a- z_l) (1-\b
z_{l})}\over{(1-\a z_{l})(z_l-\b)}}\right ]^\h-\prd\left[{{(\a-
z_l)(1-\b z_{l})}\over{(1-\a z_{l})(z_l-\b)}}\right
]^\nu\right\}\label{del}
\eeaa
is  finite   for $\a=1$
$$ \Delta
a^\n_k=\sqrt{2}\bigl({\sc\h}-\nu\bigr)\int\limits_{1-\ep}^1\prd{{
dz_l}\over{(1-z_lz_{l+1} )}}\sum_{l=1}^{2k}(1-z_l)+\const=\const.
$$
Hence  a ratio of the determinants
(\ref{cor3}) and (\ref{cn1}) is also finite  and it is not equal zero
\be
 \g \equiv
{{G(\infty)}\over{G^\n({\infty})}}={{|\cos^2\pi\nu+\sin^2\pi\nu\hv
\cdot\hv^T|}\over{|\hv^\n\cdot\hv^{\n\,\,T}|}}\mathop{=}_{(ma)\to 0}
\const.
\label{rel}
\ee
Therefore the singular over $(ma)$  factor in
$G(\infty)$ coincides with one in the determinant (\ref{cn1}).
Since the matrix  $\hv^\n$  is a Toeplitz matrix
we can calculate  (\ref{cor3}) in limit $r\to\infty,$ ${(ma)\to0}$.

Note that we have considered
the ratio (\ref{rel}) in limit
 $r=\infty$, $(ma)\to 0$.
In the paper \cite{LZ} the correlation function of the
exponential field was normalized on the correlation function
 in the massless fermion field theory.
Therefore for comparison of our results with
\cite{LZ} it is necessary to evaluate  the ratio
  (\ref{rel}) in limit $(ma)=0$,
$r\to\infty$: in other words we must verify  that the value of
 $\g$ does not depend on a sequence of the limits.

It can show that at
 $(ma)=0$ the difference of the coefficients
 (\ref{ko1}) and (\ref{ko7})  differs from (\ref{del}) on some term
decreasing with increase of $r$
 $$
 a_k -a_k^{\n}=\Delta a_k +{{\phi_k(r)}\over{r^2}}.
$$
For   $\phi_k(r)$ we have the following asymptotic evaluation
 \be
\phi_k(r)\simeq{{\ln^k r}\over{k!}}.\label{as1}\ee
As a result we get at
 $ma=0$ and $r\to\infty$
 \be
{G(r)\over
G^\n(r)}=\g\exp\biggl[{1\over{r^2}}\sum_{k=1}^{\infty}\phi_k(r)\biggr].
\label{rel1}\ee
Using  (\ref{as1}), it is not hard to sum the series in the exponent of
the right-hand side of  (\ref{rel1})  and to obtain the following
estimate
\be
{{G(r)}\over{G^\n(r)}}=\g+\oo(1/r).\label{rel2}
\ee

\renewcommand{\theequation}{4.\arabic{equation}}
\setcounter{equation}{0}

\bigskip
\centerline {\bf 4. Asymptotics of the correlation function}

\bigskip

In this section we calculate the  large distance behaviour
of the correlation function
$G(r)$. We use the determinant representation
(\ref{cor4}) for it  in this case.
Since in (\ref{cor4}) the matrix
$V^{(\nu)}_{x,x^\prime}$ is a matrix of the Toeplitz type
it can apply  the McCoy and Wu techniques \cite{MW},
which they used for the asymptotic evaluation
of the correlation function in the Ising model $(\nu=\h)$.
 This techniques requires some generalization
for $0<\nu<1$.
In Appendix B we obtained the following asymptotic expression
for the determinant of the matrix $\widehat V^{(\nu)}$
(the formula
\ref{f1}):
\be \ln |\widehat V^{(\nu)}|\mathop{=}_{r\to
\infty}(r+1)\oint{{dz}\over {z}}\ln V^{(\nu)}-\oint dz \ln P(z)
 {\partial\over{\partial z}}\ln Q(z^{-1}).  \label{f2} \ee
Substituting here
the explicit expressions for $V^(\nu)(z)$, $P(z)$ and  $Q(z)$:
\be
V^{(\nu)}(z)=P(z)Q(z^{-1}),\quad P(z)=\l{{1-\a z}\over{1-\b z}}
\r^\nu,\,\,
Q(z)=\l{{1-\b z}\over{1-\a z}} \r^\nu= {1\over{P(z)}},\label{pq1}
\ee
one gets
$$
\oint{{dz}\over z}\ln  V^{(\nu)}= \nu\l  \oint{{dz}\over z}\ln P(z)-
\oint{{dz}\over z}\ln P(z^{-1})\r =\nu\l P(0)-P(0)\r=0,
$$
$$
-\oint dz \ln P(z) {\partial\over{\partial z}}
\ln Q(z^{-1})= \nu^2 \oint dz \ln P(z)\l
{1\over{z-\a}}-{1\over{z-\b}}\r=\mu^2\l \ln P(\a)-\ln P(\b)\r.
$$
Thus
\be
|\widehat V^{(\nu)}|\mathop{=}_{r\to\infty}
\l{{P(\a)}\over{P(\b)}}\r^{\nu^2}
= \left[{{(1-\a^2)(1-\b^2)}\over{(1-\a \b)}}
\right] ^{\nu^2}.
\label{pas3}
\ee

Note that this expresion is derived for  finite values of
  $(ma)$ and $r$.
The derivation of  (\ref{f2})  is also correct
in the scaling region:  $(ma)\to 0$, $r\to\infty$, $r(ma)=\mbox{const}$.
However for $(ma)=0$, $r=\mbox{const}$ the integrals in
the right-hand side  of
(\ref{f2}) are divergent and the calculation of $|\hat V^{(\nu)}|$ at
 $(ma)=0$ and $r\to\infty$  requires the special consideration.

Note
that the following  ratio of  determinants
\be{{|\hv^{(\nu)}
(\b)|}\over{|\hv^{(\nu)}(0)|}}\mathop{=}_{r\to \infty}\left
[{{1-\b^2}\over
{(1-\a\b)^2}}\right]^{\nu^2}\mathop{=}_{(ma)=0}2^{\nu^2/2}.
\label{rat1}
\ee
is finite  both  in the scaling
regime and at the critical point.
Here we explicitly indicated a dependent of the matrix
$\hv^{(\nu)}$ on the parameter $\b$
(the matrix $\hv^{(\nu)}(0)$ is determined by the kernel
(\ref{kon})  at
 $\b=0$).

The relation
 (\ref{rat1}) it  is not hard to obtain by means of (\ref{f2}).
Really, for the ratio of the determinants in the left-hand side of
(\ref{rat1}) we get
\beaa
 \ln \l
{{|\hv^{(\nu)}(\b)|}\over{|\hv^{(\nu)}(0)|}} \r \!\!&=&\!\!
-\oint dz \ln
 P(z) {\partial\over{\partial z}}\ln Q(z^{-1})+ \nonumber\\ &&
+\oint dz
\left[\ln P(z)+\nu\ln(1-\b z)\right]{\partial\over{\partial z}}
\left[\ln
Q(z^{-1})-\nu\ln(1-\b z^{-1})\right]= \nonumber\\ \!\!&=&\!\!
-\nu\ln
  P(\b)+\nu\ln  P(0)-\nu^2 \ln (1-\b^2)+ \nonumber\\ &&+
\nu\oint dz \ln
(1-\b z) {\partial\over{\partial z}}\ln Q(z^{-1}).\label{rat2}
\eeaa
The last term in the right-hand side of
(\ref{rat2}) one integrates by parts
$$
\nu\oint dz
\ln (1-\b z){\partial\over{\partial z}}\ln Q(z^{-1})= \nu\b\oint
{{dz}\over{1-\b z}}\ln Q(z)=\nu\ln Q(\b). $$
As a result  we have
$$ \ln \l
{{|\hv^{(\nu)}(\b)|}\over{|\hv^{(\nu)}(0)|}} \r=-\nu\ln P(\b)+
\nu\ln Q
(\b)-\nu^2\ln(1-\b^2)=\ln\left[{{1-\b^2}\over{(1-\a\b)^2}}
\right]^{\nu^2}$$
 For $(ma)=0\,\,(\a=1)$ matrix elements of the matrix
$$
V^{(\nu)}_{x, \px}(0)=\oint {{dz}\over{
z}}z^{-(x-\px)}V^{(\nu)}(z)\biggm|_{\beta=0}
$$
  are calculated in the explicit form
\be V^{(\nu)}_{x, \px}(0)= {{\sin\pi\nu}\over
\pi}{1\over{\nu-x+\px}}\label{mat1}
\ee
Call to mind that $x,\px$  take the values
 $0,\ldots, r-1$.  To simplify notations  let introduce
\be
 {\hat U}_r \equiv V^{(\nu)}_{x,
\px}(0),\label{mat2}
\ee
where  the lower index for the matrix
(\ref{mat1})  indicates  its dimension.
Multiplying the matrix
 (\ref{mat1}) from the left by the  triangular matrix
 $\hl_r$ and the right  by
the  triangular matrix $\hr_r$, we get the following relation
\be
\hl_r\cdot\hu_r\cdot\hr_r=\left( \begin{array}{cc}{{\sin\pi\nu}\over
\pi}& \hat 0\\\hat 0& -\hu_{r-1}\end{array}\right),\label{mat3}
\ee
where
 \be
\hl_r= \left(
\begin{array}{cccc} {1}&{0}&\ldots&{0}\\
{-{1\over 1}}&{\nu\over 1}-1&\ldots&{0}\\
\vdots&\vdots&\ddots&\vdots\\
-{1\over{r-1}}&{0}&\ldots&{\nu\over{r-1}}-1
\end{array} \right),
\label{mat4}
\quad
\hr_r=
\left(
\begin{array}{cccc}
{\nu}&{-{1\over 1}}&\ldots&-{1\over{r-1}}\\
{0}&{\nu\over 1}+1&\ldots&{0}\\
\vdots&\vdots&\ddots&\vdots\\
0&{0}&\ldots&{\nu\over{r-1}}+1
\end{array} \right).
\label{mat5}
\ee
The determinant of a product of these matrises it is not hard
to calculate
\be
|\hl_r\cdot\hr_r|={{(-1)^{r-1}\sin\pi\nu}\over{\pi}}\cdot
{{\G(r+\nu)\G(r-\nu)}\over{\G^2(r)}}.\label{det1}
\ee
This allows to get
from
(\ref{mat3}) and (\ref{det1})
the simple recursion relation for
the determinants
$$
|\hu_{r-1}|= |\hu_r|{{\G(r+\nu)\G(r-\nu)}\over{\G^2(r)}},$$
which has the solution
 \be
|\hu_r|=\prod_{k=1}^{r}{{\G^2(k)}\over{\G(k+\nu)\G(k-\nu)}}.
\label{det2}
\ee
Using the following representation for
 $\G$-function
$$
\ln\G(z)=\int\limits_{0}^{\infty}{{dt}\over t}
\left( {{\e^{zt}-\e^{-t}}\over{1-\e^{-t}}}+(z-1)\e^{-t}\right),$$
the solution  (\ref {det2}) can be represented in the form
\be
\ln|\hu_r|=- \int\limits_{0}^{\infty}{{dt}\over t}
(1-\e^{-tr}){{\sh^2{{\nu t}\over 2}}\over{\sh^2{t\over 2}}}.
\label{det3}
\ee
For $r\to\infty$  the integral in the right-hand side of
(\ref{det3}) is  divergent for the zero.
Rewriting
(\ref{det3}) in the form
$$
\ln|\hu_r|=-\nu^2\ln r-\int\limits_{0}^{\infty}{{dt}\over t}
\left({{\sh^2{{\nu t}\over
 2}}\over{\sh^2{t\over 2}}}-\nu^2\e^{-t}\right)+\int
\limits_{0}^{\infty}{{dt}\over t}
\e^{-tr}\left({{\sh^2{{\nu t}\over 2}}\over{\sh^2{t\over 2}}}
-\nu^2\right),
$$
we can isolate this divergence.
Now the integral depending on $r$ is well defined  and we get
\be
\ln|\hu_r|\mathop{=}_{r\to \infty} -\nu^2\ln r
-\int\limits_{0}^{\infty}{{dt}\over t}
\left({{\sh^2{{\nu t}\over 2}}\over{\sh^2{t\over 2}}}-\nu^2\e^{-t}
\right)+
\mbox{O}(1/r^2).\label{det4}\ee
Taking into account
 (\ref{cor4}), (\ref{pas3}), (\ref{rat1}), (\ref{mat2}) and
(\ref{det4}),
$G(r)$
one finds
the following asymptotic evaluations for the correlation function
\be
G(r)\mathop{=}_{r\to \infty,\,\a<1}\left[{{(1-\a^2)(1-\b^2)}
\over{(1-\a \b)^2}}
\right] ^{2\nu^2}= (2\sqrt 2 ma)^{2\nu^2},\label{asy2}\ee
\be
G(r)\mathop{=}_{r\to \infty,\,\a=1}\left[{{(1-\b^2)}
\over{r(1-\a \b)^2}}\right]
^{2\nu^2}\exp\left[-2
\int\limits_{0}^{\infty}{{dt}
\over t}\left({{\sh^2{{\nu t}\over
2}}\over{\sh^2{t\over 2}}}-\nu^2\e^{-t}\right)\right]
=\frac{2^{\nu^2}\e^{-2\rho(\nu)}}{r^{2\nu^2}},
\label{asy3}
\ee
where
$$
\rho(\nu)=
\int\limits_{0}^{\infty}{{dt}
\over t}\left({{\sh^2{{\nu t}\over
2}}\over{\sh^2{t\over 2}}}-\nu^2\e^{-t}\right),\quad
 2(ma)\mathop{=}_{\a\to 1}1-\alpha^2.
$$
Using these asymptotics one can assume that in the scaling limit
$(r\to \infty,\, ma\to 0, \,mar = s= \const)$
the correlation function $G(r)$
has the scaling form
\be
F(s,\nu)=\lim r^{2\nu^2}G(r)=
2^{3\nu^2}s^{2\nu^2}f(s,\nu),
\label{sc1}
\ee
where the  function  $f(s,\nu)$
has the asymptotics
\be
f(s,\nu)=\left\{\begin{array}{rcl}&1
\quad \mbox{for} \,\,s\to\infty,\\
&2^{-2\nu^2}\e^{-2\rho(\nu)}s^{-2\nu^2}\quad\mbox{for}
\,\,s\to 0.\\
\end{array}\right.
\label{sc2}
\ee
This scaling behaviour follows from the  arquments:

1) the two-point scaling function $F(s,\nu)$  have to interpolate
between the behaviour at large distance away from the critical temperature
$(r\to \infty,\, ma\to 0,\,\, \mbox{so that}\,\, s\to \infty)$
and the behaviour at large distance at the critical point
$(r\to \infty,\, ma\to 0,\,\, \mbox{so that}\,\, s\to 0)$,
that is the asymptotics (\ref{sc2}) have to result in asymptotics
(\ref{asy2}), (\ref{asy3}).

2) for $\nu=\h$ the scaling function (\ref{sc1})  and the asymptotics
(\ref{sc2}) have to coincide with square of
the scaling function  and the
asymptotics for  the two-point correlation function of
the disorder field in
the Ising model \cite{MC1, MC2, T1}.

It is convenient to define the ratio $h(Rm)$, which does not depend
 on a normalization and
the  lattice cutoff $a$,
\be
h(Rm)= {{G(R)|_{R\to \infty}}
\over{G(R)|_{R\to 0}}}={{F(s,\nu)|_{s\to \infty}}
\over{F(s,\nu)|_{s\to 0}}}
=(2mR)^{2\nu^2}\exp\left[2\int\limits_{0}
^{\infty}{{dt}\over t}\left({{\sh^2{{\nu t}\over 2}}
\over{\sh^2{t\over 2}}}
-\nu^2\e^{-t}\right)\right],\label{asy4}\ee
where $R=(ra)$  denotes a dimensional distance.

To compare this asymptotics with the result of the paper
\cite{LZ} in the free fermion point we
use  the following normalization for the correlation function
for $R\to 0$ \cite{LZ}
\be
G(R)={{R}^{-2\nu^2}}.
\label{bezm}
\ee
Using this normalization, from the ratio
  (\ref{asy4})  we obtain
\be
G(R) \mathop{=}_{R\to \infty,\,m\neq 0}
\langle \mu_\nu\rangle ^2=
(2m)^{2\nu^2}\exp\left[2\int\limits_{0}
^{\infty}{{dt}\over t}\left({{\sh^2{{\nu t}\over 2}}
\over{\sh^2{t\over 2}}}
-\nu^2\e^{-t}\right)\right].\label{asy5}
\ee

\renewcommand{\theequation}{5.\arabic{equation}}
\setcounter{equation}{0}

\bigskip
\centerline {\bf 5. Conclusion}

\bigskip

In this work a determinant representation for
the two-point correlation
function of the twist field is obtained in the
lattice fermion field model.
 The large distance behaviour of  this correlation function
is calculated
and the vacuum expectation values of the twist fields  is found.
This value  differs from
the vacuum expectation value of the exponential field
by the constant.

For explanation of this fact one can bring the following
arguments.
It is known that the quantum fields defined by the commutation relations
(\ref{ex}) realize the operator solution of the isomodronomic deformation
problem for the  Dirac equation \cite{SJM}.
The correlation function of the twist fields (\ref{cor})
one can interpret as the functional integral in the Dirac
fermion theory with
given monodromy properties for fermion fields.
Then this correlation function can be connected  with other solution
of the isomodronomic deformation problem for the Dirac equation.
In order to check this asssumption it is necessary  to calculate the
determinant (\ref{cor3}) and to obtain  a differential equation for the
correlation function of the twist fields.

Note that with  our point of  view in Appendix B of the paper \cite{LZ}
instead of the vacuum expectation values of the exponential fields
 the vacuum expectation values of the twist fields
was calculated
but since   the normalization on behaviour of the correlation function of
exponential field for $r\to 0$ was used the right result
(\ref{asy6}) for
 the vacuum expectation values of the exponential fields
was obtained.

The authors thank N.~Slavnov,  S.~Pakuliak,
S.~Lukyanov and M.~Lashkevich  for some useful
discussions.   One of us (V.Sh.) thanks A.~Morozov
for his hospitality at the Institute of Theoretical and
Experimental Physics
   and Prof. G. von Gehlen for his hospitality
at Physikalisches Institute d. Universit\"at Bonn.

This work was perfomed with financial support from
Ukrainian Found Fundamental Reseachers (project No. 2.5.1/051)
and the INTAS program (Grant No. INTAS-97-1312).

\renewcommand{\theequation}{A.\arabic{equation}}
\setcounter{equation}{0}

\bigskip

\centerline {\bf 6. Appendix A}

\bigskip

In this Appendix we obtain  the  following representation
for the determinant in (\ref{cor4})  for $r\to\infty$
\be
g(r)=|\cos^2\pi\nu +\sin^2\pi\nu \hc\cdot\hc^T|\mathop{=}_{r\to\infty}
|1-\sin^2\pi\nu\ha|^2
%\cdot |1-\sin^2\pi\nu\hb|,
\label {pd1}
\ee
where for convenience we introduced the matrix
\be
C_{xy}=
{1\over{2\pi i}}\oint_{|z|=1}{{dz}\over{z}}
z^{-(x-y)}C(z),\quad
C(z)=\left[{{(1-\a z)(1-\b z^{-1})}\over{(1-\a z^{-1})(1-\b z)}}
\right] ^\mu,
\label{pm1}
\ee
It has dimension
$(r+1)\times(r+1)$ ($x,y=0,1,\ldots,r$)  and $0<\mu<1$.
For $\mu=\h$ this matrix coincides with the matrix
$V_{x, y}$  in the determinant representation
of the correlation function
(\ref{cor3}).  In (\ref{pd1}) the matrix $\ha$ has the form
$$ A_{x,y}
=\oint\limits_{|z_1z_2|<1} {{dz_1dz_2 (z_1)^x (z_2)^{y}}\over{
(1-z_1z_2)}} C^{-1}(z_1)C^{-1}(z_2). $$

Let us represent the product
$\hc\cdot\hc^T$  in the following form
$$
(\hc\cdot\hc^T)_{xy}=
\oint_{|z|<|\xi|<1}
{{d(z\xi)}\over{z\xi}}z^{-x}\xi^yC(z)C(\xi^{-1})
\sum_{\px=0}^{r}
\left( {z\over\xi}\right)^{\px}=
$$
$$
\oint_{|\xi|<1<|z|}
{{d(z\xi)z^{-x}\xi^y}\over{z\xi(1-z/\xi)}}{{C(z)}\over{C(\xi)}}
+
\oint
{{dz}\over{z}}z^{-x+y}
-
\oint_{|z|<|\xi|<1}
{{d(z\xi)z^{r+1-x}\xi^{y-r-1}}\over{z\xi(1-z/\xi)}}{{C(z)}
\over{C(\xi)}}=
$$
 \be
\delta(x-y)-A_{xy}-B_{xy},
\label{pm2}
\ee
where
$$
A_{xy}=-
\oint_{|\xi|<1<|z|}
{{d(z\xi)z^{-x}\xi^y}\over{z\xi(1-z/\xi)}}{{C(z)}\over{C(\xi)}}
\mathop{=}_{z\to 1/z}
\oint
{{d(z\xi)z^{x}\xi^y}\over{1-z\xi}}{1\over{C(z)C(\xi)}}
$$
 $$
B_{xy}=
\oint_{|z|<|\xi|<1}
{{d(z\xi)z^{r+1-x}\xi^{y-r-1}}\over{z\xi(1-z/\xi)}}
{{C(z)}\over{C(\xi)}}
\mathop{=}_{\xi\to 1/\xi}
\oint
{{d(z\xi)z^{r+1-x}\xi^{y-r-1}}\over{1-z\xi}}C(z)C(\xi)
$$
In (\ref{pm2}) we summed a geometric progression over
 $\px$  and used $C(z^{-1})=C^{-1}(z)$.
Substituting (\ref{pm2}) in (\ref{pd1}), we get
\be
g(r)=|1-\sin^2\pi\nu(\ha+\hb)|=|(1-s^2\ha)\cdot(1-s^2\hb)
-s^4\ha\cdot\hb|,
\label{pd2}
\ee
where $s=\sin\pi\nu$.

Note that matrix elements of the matrices
 $\ha$ and  $\hb$ have the following asymptotic behaviour
$$
A_{xy}\mathop{\sim}_{x+y\gg 1} \a^{x+y},
\quad
B_{xy}\mathop{\sim}_{2r-x-y\gg 1} \a^{2r-x-y}.
$$
and therefore  the product
 $\ha\cdot\hb$ is a exponential small matrix.
Really,
$$
(\ha\cdot\hb)_{xy}=
\oint {{d(z_1...z_4)}\over{(1-z_1z_2)(1-z_3z_4)}}
{{C(z_3)C(z_4)}\over{C(z_1)C(z_2)}}(z_1)^x (z_4)^{r-y}
\sum_{\px=0}^{r}(z_2)^{\px} (z_3)^{r-\px}=
$$
$$
\oint {{d(z_1...z_4)}\over{(1-z_1z_2)(1-z_3z_4)}}
{{C(z_3)C(z_4)}\over{C(z_1)C(z_2)}}(z_1)^x (z_4)^{r-y}
{{(z_3)^{r+1}-(z_2)^{r+1}}\over{z_3(1-z_2/z_3)}}
\mathop{\sim}_{r\to\infty}
$$
\be
{{\a^r\ln r}\over{r}}
\left(
\oint {{dz z^x}\over{1-\a z}}C(z^{-1})
\right)
\left(
\oint {{dz z^{r-y}}\over{1-\a z}}C(z)
\right)
\label{pm3}
\ee
Taking into account (\ref{pm3}) in  (\ref{pd2}),
we obtain at
$r\to\infty$
\be g(r) \mathop{=}_{r\to\infty} |(1-s^2\ha)|\cdot|(1-s^2\hb)|.
\label{pd3}
\ee
Let us show that $|(1-s^2\ha)|=|(1-s^2\hb)|$ for  $r\to\infty$.
For this we write the matrix
 $\hb$ in the form
$$
\hb=\hj\cdot\hd\cdot\hj,
$$
where
$$
J_{xy}=\delta(x+y-r)=
\left(
\begin{array}{cccc}
{0}&{\ldots}&0&{1}\\
{0}&\ldots&1&{0}\\
\vdots&\vdots&\ddots&\vdots\\
{1}&\ldots&{0}&0
\end{array} \right), \quad (\hj\cdot\hj)_{xy}=\delta(x-y),
$$
\be
D_{xy} =
\oint
{{d(z\xi)z^{x}\xi^{y}}\over{1-z\xi}}C(z)C(\xi).
\label{pm4}
\ee

Using this representation, it is not hard to show that
\be
|(1-s^2\ha)|=|(1-s^2\hd)|
\label{pd4}
\ee
The determinants in
 (\ref{pd4}) it is convenient to decompose  in the  series
$$
\ln|1-s^2\ha|=\mbox{Sp}\ln\left( 1-s^2\ha\right)=-
\sum_{k=1}^{\infty}{{s^{2k}}\over{k}}a_{2k},
$$
$$
\ln|1-s^2\hd|=\mbox{Sp}\ln\left( 1-s^2\hd\right)=-
\sum_{k=1}^{\infty}{{s^{2k}}\over{k}}d_{2k},
$$
where
$$
a_{2k}=\mbox{Sp}\left( \ha\right)^k, \quad
d_{2k}=\mbox{Sp}\left( \hd\right)^k.
$$
For  $r\to\infty$ the matrix $\ha$  and  $\hd$
can be represented  in the factorized form
$$
\ha\mathop{=}_{r\to\infty}  \hf^{(a)}\cdot\hf^{(a)}, \quad
\hd
\mathop{=}_{r\to\infty}
\hf^{(d)}\cdot\hf^{(d)}, \quad F_{xy}^{(a)}=\oint dz z^{x+y}C(z),\quad
F_{xy}^{(d)}=\oint dz z^{x+y} C^{-1}(z),
$$
so that
$$
a_{2k}=\l\mbox{Sp}{\hf^{(a)}}\r^{2k}=\oint\prod_{l=1}^{2k}
\l {{dz_lC^{-1}(z_l)}\over{1-z_lz_{l+1}}}\r, \quad z_{2k+1}
\equiv z_1,
$$
\be
d_{2k}=\l\mbox{Sp}{\hf^{(d)}}\r^{2k}=\oint\prod_{l=1}^{2k}
\l {{dz_lC(z_l)}\over{1-z_lz_{l+1}}}\r, \quad z_{2k+1}\equiv z_1.
\label{pk1}
\ee
Consider, for example, the case
 $k=1$
$$
d_{2}=
\oint_{|z_1z_2|<1}{{dz_1dz_2}\over{(1-z_1z_2)^2}}C(z_1)C(z_2)
\mathop{=}_{z_2\to1/z_2}
\oint_{|z_1|<|z_2|}{{dz_1dz_2}\over{(z_1-z_2)^2}}{{C(z_1)}
\over{C(z_2)}}=
$$
\be
\oint_{|z_2|<|z_1|}{{dz_1dz_2}\over{(z_1-z_2)^2}}{{C(z_1)}
\over{C(z_2)}}+
\oint dz C(z){{\partial\,}\over{\partial z}} C^{-1}(z).
\label{pk2}
\ee
Such as in the right-hand side of
 (\ref{pk2})
the second term is a integral over a closed contour from a derivative
it is equal to  zero
$$
\oint dz C(z){{\partial\,}\over{\partial z}} C^{-1}(z)=
-\oint dz {{\partial\,}\over{\partial z}} \ln C(z)=0.
$$
In the first term in the  right-hand side of
 (\ref{pk2}) we did the  replacement
 $z_1\to (z_1)^{-1}$
\be
d_2=
\oint_{|z_1z_2|<1}{{dz_1dz_2}\over{(1-z_1z_2)^2}}{1
\over{C(z_1)C(z_2)}}=
a_2.
\label{pk3}
\ee
For $k>1$ it is not hard to get the recurent relation
\be
d_{2k}=a_{2k}+d_{2k-2}-a_{2k-2}.
\label{rek}
\ee

For this in the integral representation
(\ref{pk1})   it is necessary
to do the  replacement
$z_l\to(z_l)^{-1}$. After that we obtain  new  integration contours
  with  $|z_lz_{l+1}|>1$. Contracting sequentially these contours
over the variables $z_l$ so that   $|z_lz_{l+1}|<1$
and calculating residues
in the poles $z_l=(z_{l+1})^{-1}$, we get (\ref{rek}).

Taking into account the "initial condition"
(\ref{pk3}), it is easy to check that a solution
of the recurent relation
 (\ref{rek}) is
$a_{2k}=d_{2k}$. From here we obtain  (\ref{pd4}).
Then
 (\ref{pd4})   and   (\ref{pd3})
lead to
 (\ref{pd1}).

Note that putting $\mu=\h$
in  (\ref{pm1}),  we get (\ref{fa1}) from (\ref{pd1}),
and  putting $\nu=h$  in (\ref{pd1}) and $\mu =\nu$  in (\ref{pm1}),
 we get   (\ref{cn2}) from  (\ref{pd1}).

\renewcommand{\theequation}{B.\arabic{equation}}
\setcounter{equation}{0}
\bigskip

\centerline {\bf 7. Appendix B}

\bigskip

In this Appendix we obtain  the  asymptotic evaluation
 of
  $|\hv^{(\nu)}|$ for $r\to \infty$
\be
 |\hv^{(\nu)}|\mathop{=}_{r\to \infty}
%(N+1)\int{{dz}\over z}\ln V^{(\nu)}(z) -\int dz
%\ln P(z){\partial\over{\partial z}}
%\ln Q(z^{-1})=
\left[{{(1-\a^2)(1-\b^2)}\over{(1-\a \b)}}
\right] ^{\nu^2}.
\label{pas1}
\ee

It is convenient to introduce  the following notations
\be
V_{xy}^{(\nu)}\equiv B_{xy}=
\oint{{dz}\over{z}} z^{-x+y}B(z),
\quad B(z)=P(z)Q(z^{-1}),
\label{pma1}
\ee
where $x,y=0,1,...,r$, the functions $P(z)$ and $Q(z)$
is anylitical in the circle
$|z|\le 1$
\be P(z)=\l{{1-\a z}\over{1-\b z}} \r^\nu,\quad Q(z)=\l{{1-\b
z}\over{1-\a z}} \r^\nu= {1\over{P(z)}},\quad \a>\b\le 0, \label{pq}
\ee
$\a$
and  $\b$  are determined in (\ref{par}).

Let us define
 \be
B_{xy}(\mu)=
\oint{{dz}\over{z}} z^{-x+y}B^\mu(z),\quad
B^\mu(z) =P^\mu(z)Q^\mu(z^{-1}),
\label{pde}
\ee
and
$$
f(\mu)=\ln |\hb(\mu)|,\quad f(0)=0,\,\, f(1) =\ln|\hb|,\,\,
B_{xy}(1)=
B_{xy},\,\,
B_{xy}(0)=\d(x-y),
$$

For calculation of
 $|\hb|$  we use the formula
\be
\ln|\hb| = f(1)=\int_{0}^{1}f^{\prime}(\mu), \quad
f^{\prime}(\mu)=\mbox{Sp}\hb^\prime(\mu)\cdot \hb^{-1}(\mu).
\label{fp}
\ee

Let us find the inverse matrix
 $\hb^{-1}(\mu)$.
Note that the matrix
 $\hb(\mu)$ for $0\le\mu\le 1$  has the same  analytic behaviour
as the matrix $\hb$, therefore,  for simplicity  we  only derive
  the inverse matrix $\hb^{-1}$.

Show that matrix $\hd$
\be
\hd= \hl - \hee
 -\hj\cdot\hee^{T}\cdot\hj
\label{pma2}
\ee
is inverse to
 $\hb$ up to an exponentially small term
$$
\hb\cdot\hd= I +\mbox{O}(\a^r) \quad \mbox{at} \,\,r\to\infty,
$$
where $J^{xy}=\d(x+y-r)$,
\be
E_{xy}=
\oint_{|z_1z_2|<1}
{{dz_1dz_2}\over{1-z_1z_2}}
{{(z_1)^x(z_2)^y}\over{P(z^{-1}_1)Q(z^{-1}_2)}}=
\oint_{|z_2|<|z_1|}{{dz_1dz_2}\over{z_1(z_1-z_2)}}
{{(z_1)^{-x}(z_2)^y}\over{P(z_1)Q(z^{-1}_2)}},
\label{e}
\ee
$$
L_{xy}=\oint {{dz}\over{z}}  {{z^{-x+y}}\over{B(z)}}=B^T_{xy},
\quad\mbox{if}\,\, B(z^{-1})=B^{-1}(z),
$$
since
$$
B^T_{xy}=B_{yx}= \oint {{dz}\over{z}}  {{z^{-y+x}}{B(z)}}
\mathop{=}_{z\to 1/z}
\oint {{dz}\over{z}}  {{z^{-x+y}}{B(z^{-1})}}
$$
and in our case
 $B(z^{-1})=B^{-1}(z)$.

At the first we calculate
 $\hb\cdot\hl$
$$
\l\hb\cdot\hl\r_{xy}=\oint_{|z_1|<|z_2|}
{{dz_1dz_2B(z_1)}\over{z_1z_2B(z_2)}}
{{1-(z_1/z_2)}^{r+1}\over{1-z_1/z_2}}
(z_1)^{-x}(z_2)^y=\d(x-y)-
$$
\be
\oint_{|z_1z_2|<1}
{{dz_1dz_2B(z^{-1}_1)}\over{z_1z_2B(z_2)}}
(z_1)^{x}(z_2)^y -
\oint_{|z_1z_2|<1}
{{dz_1dz_2B(z_1)}\over{z_1z_2B(z^{-1}_2)}}
(z_1)^{r-x}(z_2)^{r-y}.
\label{bl}
\ee
Now one calculates the product
 $\hb\cdot\hee$
$$
\l\hb\cdot\hee\r_{xy}=
\oint_{|z_1|<|z_3|<|z_2|}
{{d(z_1z_3)(z_1)^{-x}(z_3)^y}\over{z_1z_2(z_2-z_3)}}
{{B(z_1)}\over{P(z_2)Q(z_3^{-1})}}
{{1-(z_1/z_2)}^{r+1}\over{1-z_1/z_2}}=
$$
$$
\oint_{|z_1|<|z_3|}
{{d(z_1z_3)(z_1)^{-x}(z_3)^y}\over{z_1(z_3-z_1)}}
{{B(z_1)}\over{B(z_3)}}+
\oint_{|z_1|<|z_3|}
{{d(z_1z_3)(z_1)^{-x}(z_3)^y}\over{z_1(z_1-z_3)}}
{{Q(z^{-1}_1)}\over{Q(z^{-1}_3)}}\,-
$$
\be
\oint_{\a<|z_i|<1}
{{d(z_1z_2z_3)(z_1)^{r-x}(z_3)^y(z_2)^{r+1}}\over{(1-z_1z_2)(1-z_2z_3)}}
{{B(z_1)}\over{P(z^{-1}_2)Q(z_3^{-1})}}.
\label{be}
\ee
In (\ref{be}) the integration over $z_2$ yields a contribution
which is propotional to $\a^{r+1}$ and therefore,
the last term in the right-hand side of (\ref{be})
 is exponentially small for any $x,y$.
The second term  in the right-hand side of (\ref{be})
gives  the following contribution at the contraction of the contour
over $z_3$
\be
\oint_{|z_3|<|z_1|}
{{d(z_1z_3)(z_1)^{-x}(z_3)^y}\over{z_1(z_1-z_3)}}
{{Q(z^{-1}_1)}\over{Q(z^{-1}_3)}}-\d(x-y)=-\d(x-y),
\label{be1}
\ee
where  we took into account that  the first term
in the left-hand side of (\ref{be1}) reduces  to  zero at  the extention
of the contour over $z_3$ to one of infinite radius.

Similarly for the first term
in the right-hand side of (\ref{be}) one gets
\be
\oint_{|z_3|<|z_1|}
{{d(z_1z_3)(z_1)^{-x}(z_3)^y}\over{z_1(z_3-z_1)}}
{{B(z_1)}\over{B(z_3)}}+\d(x-y).
\label{be2}
\ee

Taking into account
 (\ref{be1}) and (\ref{be2}),  we obtain
$$
\l\hb\cdot\hee\r_{xy}=
-\oint_{|z_2|<|z_1|}
{{d(z_1z_2)(z_1)^{-x}(z_2)^y}\over{z_1(z_2-z_1)}}
{{B(z_1)}\over{B(z_2)}}-
$$
$$
\oint_{\a<|z_i|<1}
{{d(z_1z_2z_3)(z_1)^{r-x}(z_3)^y(z_2)^{r+1}}\over{(1-z_1z_2)(1-z_2z_3)}}
{{B(z_1)}\over{P(z^{-1}_2)Q(z_3^{-1})}}=
$$
\be
\mathop{=}_{z_1\to 1/z_1}
-\oint_{|z_1z_2|<1}
{{d(z_1z_2)(z_1)^{x}(z_2)^y}\over{(1-z_1z_2)}}
{{B(z^{-1}_1)}\over{B(z_2)}}
- \mbox{O}(\a^{r}).
\label{be3}
\ee
From here it follows that
$\hb\cdot\hee$  coincides up to $\mbox{O}(\a^{r})$
with the
second term in the right-hand side of (\ref{bl})  of the product
$\hb\cdot\hl$.

Now let us consider the product
$\hb\cdot\hj\hee^T\hj$
\be
\hb\cdot\hj\hee^T\hj=\hj\cdot\l\hj\cdot\hb\cdot\hj\cdot\hee^T\r\cdot\hj=
0\hj\cdot\l\hb^T\cdot\hee^T\r\cdot\hj,
\label{be4}
\ee
where we used
 $\hj\cdot\hb\cdot\hj=\hb^T$.

A transition from
 $\hb$, $\hee$  to $\hb^T$, $\hee^T$ means the following transformations
in contour integrals
(\ref{pma1}) and  (\ref{e}): $B(z)\to B(z^{-1})$, $P(z)\leftrightarrow
Q(z)$
and therefore, no calculating, we have instead of
(\ref{be3})
$$
\l\hb^T\cdot\hee^T\r_{xy}=
-\oint_{|z_1z_2|<1}
{{d(z_1z_2)(z_1)^{x}(z_2)^y}\over{(1-z_1z_2)}}
{{B(z_1)}\over{B(z^{-1}_2)}}-
$$
\be
\oint_{\a<|z_i|<1}
{{d(z_1z_2z_3)(z_1)^{r-x}(z_3)^y(z_2)^{r+1}}\over{(1-z_1z_2)(1-z_2z_3)}}
{{B(z^{-1}_1)}\over{Q(z^{-1}_2)P(z_3^{-1})}}
\label{be5}
\ee
and from
   (\ref{be4}) and (\ref{be5})  one gets
\be
\hb\cdot\hj\hee^T\hj
=
-\oint_{|z_1z_2|<1}
{{d(z_1z_2)(z_1)^{r-x}(z_2)^{r-y}}\over{(1-z_1z_2)}}
{{B(z_1)}\over{B(z^{-1}_2)}}
- \mbox{O}(\a^{r}).
\label{be6}
\ee
Collecting together
 (\ref{bl}),  (\ref{be3}) and (\ref{be6}), one proves that
$\hd=\hb^{-1}+ \mbox{O}(\a^{r})$.
Thus  for
  $f^{\prime}(\mu)$ in (\ref{fp}) we obtain
\be
f^{\prime}(\mu)=\mbox{Sp}\hb^\prime(\mu)\cdot\left[
\hl(\mu)-\hee(\mu)-\hj\cdot\hee^T(\mu)\cdot \hj
\right].
 \label{fp1}
\ee
Note that
$$
\mbox{Sp}\hb^\prime\cdot\hj\cdot\hee^T\cdot \hj =
\mbox{Sp}\hj\cdot\hb^\prime\cdot\hj\cdot\hee^T =
\mbox{Sp}\left(\hb^\prime\right)^T\cdot\hee^T =
\mbox{Sp}\left(\hb^\prime\right)^T\cdot\hee^T =
\mbox{Sp}\hb^\prime(\mu)T\cdot\hee(\mu).
$$
From here one gets
$$
f^{\prime}(\mu)=\mbox{Sp}\hb^\prime(\mu)\left(\hl(\mu)-2\hee(\mu)\right).
$$
Let us denote by
\be
U_1=\mbox{Sp}\left(\hb^\prime(\mu)\hl(\mu)\right),\quad
U_2=\mbox{Sp}\left(\hb^\prime(\mu)\hee(\mu)\right),\quad
f^{\prime}(\mu)= U_1-2U_2.
\label{fp3}
\ee
At the first we calculate
 $U_1$
$$
U_1=\sum_{x,y}\oint{{dz_1dz_2}\over {z_1z_2}} z_1^{-y+x}B^{\mu}(z_1)
\ln B(z_1) z_2^{-x+y}B^{-\mu}(z_2)=
$$
$$
\oint_{|z_1|<|z_2|}{{dz_1dz_2}\over {z_1z_2}}
{{B^{\mu}(z_1)\ln B(z_1)}
\over{B^{\mu}(z_2)}}
{{[1-(z_1/z_2)^{r+1}][1-(z_2/z_1)^{r+1}]}\over{(1-z_1/z_2)(1-z_2/z_1)}}=
$$
$$
(r+1)\oint
{{dz}\over {z}}\ln B(z)-
\l\oint_{|z_1|<|z_2|}+\oint_{|z_2|<|z_1|} \r
{{dz_1dz_2}\over {(z_1-z_2)^2}}
{{B^{\mu}(z_1)\ln B(z_1)}
\over{B^{\mu}(z_2)}}
+
$$
$$
\oint_{|z_1|<|z_2|}
{{dz_1dz_2}\over {(z_1-z_2)^2}}
{{B^{\mu}(z_1)\ln B(z_1)}
\over{B^{\mu}(z_2)}}\l{{z_1}\over{z_2}}\r^{r+1}
+
\oint_{|z_2|<|z_1|}
{{dz_1dz_2}\over {(z_1-z_2)^2}}
{{B^{\mu}(z_1)\ln B(z_1)}
\over{B^{\mu}(z_2)}}\l{{z_2}\over{z_1}}\r^{r+1}.
$$
In this expression the last two terms  are exponentially
small ($\sim \mbox{O}(\a^{r}))$) and therefore, \be U_1= (r+1)\oint
{{dz}\over {z}}\ln B(z)-
\l\oint_{|z_1|<|z_2|}+\oint_{|z_2|<|z_1|} \r
{{dz_1dz_2}\over {(z_1-z_2)^2}}
{{B^{\mu}(z_1)\ln B(z_1)}
\over{B^{\mu}(z_2)}}+\mbox{O}(\a^{r})
\label{u1}
\ee
Now let us calculate
 $U_2$
$$
U_2=\sum_{x,y}
\oint{{d(z_1z_2z_3}\over {z_1(1-z_2z_3)}}
 z_1^{-y+x}z_2^{y}z_3^{x}{{B^{\mu}(z_1)
\ln B(z_1)}\over{P^\mu(z^{-1}_2)Q^\mu(z_3^{-1}
)}} =
$$
$$
\oint_{|z_3|<|z_2|<|z_1|<1}{{d(z_1z_2z_3}\over {z_1(1-z_2z_3)}}
{{B^{\mu}(z_1)
\ln B(z_1)}\over{P^\mu(z^{-1}_2)Q^\mu(z_3^{-1})}}
{{[1-(z_1z_2)^{r+1}][1-(z_3/z_1)^{r+1}]}\over{(1-z_1z_2)(1-z_3/z_1)}}=
$$
\be
\oint_{|z_3|<|z_1|<|z_2|}{{d(z_1z_2z_3}\over {(z_1-z_3)(z_2-z_1)
(z_2-z_3)}}
{{B^{\mu}(z_1)
\ln B(z_1)}\over{P^\mu(z_2)Q^\mu(z_3^{-1}])}}
+ \mbox{O}(\a^{r}).
\label{u2}
\ee
In the right-hand side of (\ref{u2})
contracting  the contour over $z_2$, we obtain
\be
\oint_{|z_3|<|z_1|}{{d(z_1z_3}\over {(z_1-z_3)^2}}
{{Q^{\mu}(z^{-1}_1)
\ln B(z_1)}\over{Q^\mu(z_3^{-1}])}} -
\oint_{|z_3|<|z_1|}{{d(z_1z_3}\over {(z_1-z_3)^2}}
{{B^{\mu}(z_1)
\ln B(z_1)}\over{B^\mu(z_3)}}
+ \mbox{O}(\a^{r}).
\label{ua2}
 \ee
In the first term of the right-hand side of
 (\ref{ua2}) let us move the contour over
$z_3$ to $\infty$. Then
$$
U_2=
\oint dz Q^\mu(z^{-1})\left[\ln P(z)+ \ln Q(z^{-1})\right]
{\partial\over{\partial z}}\l {1\over{Q^\mu(z^{-1})}}\r=
$$
\be
\mu \oint dz \ln P(z) {\partial\over{\partial z}}
\ln Q(z^{-1})
+ \mbox{O}(\a^{r}).
\label{ua3}
\ee
Substituting
(\ref{u1}) and (\ref{ua3}) in    (\ref{fp3}),
one gets
$$
f^{\prime}(\mu)=
(r+1)\oint
{{dz}\over {z}}\ln B(z)-
\l\oint_{|z_1|<|z_2|}-\oint_{|z_2|<|z_1|} \r
{{dz_1dz_2}\over {(z_1-z_2)^2}}
{{B^{\mu}(z_1)\ln B(z_1)}
\over{B^{\mu}(z_2)}}-
$$
\be
-2\mu \oint dz \ln P(z) {\partial\over{\partial z}}
\ln Q(z^{-1})+\mbox{O}(\a^{r}).
\label{fp4}
\ee
The difference of the integrals over
 $z_2$  in the second term of the right-hand side of
 (\ref{fp4}) is a integral over
 $z_2$  around the point $z_1$, that is
$$
\oint_{(z_1)}{{dz_2}\over {(z_1-z_2)^2}}
{{1}
\over{B^{\mu}(z_2)}}=
{\partial\over{\partial z}}
\l{1\over{B^\mu(z)}}\r_{z=z_1}.
$$
Then the second term in the right-hand side of
 (\ref{fp4}) has the form
 $$
\mu \oint dz \ln B(z) {\partial\over{\partial z}}
\ln B(z)= {\mu\over 2}\int dz
{\partial\over{\partial z}}
\l\ln^2 B(z)\r =0,
$$
since the integral over a closed contour from a total derivative
is equal to  the zero.

Thus
\be
f^{\prime}(\mu)=
(r+1)\oint
{{dz}\over {z}}\ln B(z)-
-2\mu \oint dz \ln P(z) {\partial\over{\partial z}}
\ln Q(z^{-1}).
\label{fp5}
\ee
Integrating this expression over
 $\mu $ in the limits $[0,1]$,
we obtain
\be
f(1)=\ln|\hb|=
(r+1)\oint
{{dz}\over {z}}\ln B(z)
-\oint dz \ln P(z) {\partial\over{\partial z}}
\ln Q(z^{-1}).
\label{f1}
\ee
Substituting here the explicit expressions for
 $B(z)$, $P(z)$ and  $Q(z)$ from
(\ref{pma1}) and (\ref{pq}) we have
$$
\oint{{dz}\over z}\ln B(z)= \nu\l  \oint{{dz}\over z}\ln P(z)-
\oint{{dz}\over z}\ln P(z^{-1})\r =\nu\l P(0)-P(0)\r=0,
$$
$$
-\oint dz \ln P(z) {\partial\over{\partial z}}
\ln Q(z^{-1})= \nu^2 \oint dz \ln P(z)\l
{1\over{z-\a}}-{1\over{z-\b}}\r=\mu^2\l \ln P(\a)-\ln P(\b)\r.
$$
Thus
\be
|\widehat V^{(\nu)}|=|\hb|\mathop{=}_{r\to\infty}
\l{{P(\a)}\over{P(\b)}}\r^{\nu^2}
= \left[{{(1-\a^2)(1-\b^2)}\over{(1-\a \b)}}
\right] ^{\nu^2}.
\label{pas2}
\ee

\bigskip
{\small
\renewcommand{\refname}{\ }

}

\end{document}